\documentstyle[emulateapj,apjfonts,psfig]{article}

\lefthead{J.~M.~Miller et al.}  
\righthead{GX 339-4 During Outburst Decline}

\received{Date}
\revised{Date}
\accepted{Date}

\journalid{vol}{date}
\articleid{1}{4}
\paperid{id}

\cpright{AAS}{1999}
\ccc{x}

\begin{document}

\title{Chandra/HETGS Spectroscopy of the Galactic Black Hole
GX~339$-$4: A Relativistic Iron Emission Line and
Evidence for a Seyfert-like Warm Absorber}

\author{J.~M.~Miller\altaffilmark{1,2}, 
        J.~Raymond\altaffilmark{1},
        A.~C.~Fabian\altaffilmark{3},
	J.~Homan\altaffilmark{4},
	M.~A.~Nowak\altaffilmark{5},
	R.~Wijnands\altaffilmark{6},
	M.~van~der~Klis\altaffilmark{7},
	T.~Belloni\altaffilmark{4},		
	J.~A.~Tomsick\altaffilmark{8},
	D.~M.~Smith\altaffilmark{9},
	P.~A.~Charles\altaffilmark{10},
	W.~H.~G.~Lewin\altaffilmark{5},	
	}

\altaffiltext{1}{Harvard-Smithsonian Center for Astrophysics, 60
	Garden Street, Cambridge, MA 02138, jmmiller@cfa.harvard.edu}
\altaffiltext{2}{NSF Astronomy and Astrophysics Fellow}
\altaffiltext{3}{Institute of Astronomy, University of Cambridge,
        Madingley Road, Cambridge CB3 OHA, UK}
\altaffiltext{4}{INAF -- Osservatorio Astronomico di Brera, Via
	E. Bianchi 46, 23807 Merate, IT}
\altaffiltext{5}{Center~for~Space~Research and Department~of~Physics,
        Massachusetts~Institute~of~Technology, Cambridge, MA
        02139--4307}
\altaffiltext{6}{School of Physics and Astronomy, University of
	St. Andrews, North Haugh, St. Andrews Fife, KY16 9SS, UK}
\altaffiltext{7}{Astronomical Institute ``Anton Pannekoek,''
        University of Amsterdam, and Center for High Energy
        Astrophysics, Kruislaan 403, 1098 SJ, Amsterdam, NL}
\altaffiltext{8}{Center for Astrophysics and Space Sciences, Code
	0424, University of California, San Diego, La Jolla, CA 92093}
\altaffiltext{9}{Santa Cruz Institute for Particle Physics, University
	of California at Santa Cruz, 1156 High Street, Santa Cruz, CA
	95064}
\altaffiltext{10}{Department of Physics and Astronomy, University of
        Southampton, SO17 1BJ, UK}

\keywords{Black hole physics -- relativity -- stars: binaries
(GX339$-$4) -- stars: binaries (XTE J1650$-$500) -- physical data and
processes: accretion disks}

\authoremail{jmmiller@cfa.harvard.edu}

\label{firstpage}

\begin{abstract}
We observed the Galactic black hole GX~339$-$4 with the {\it Chandra}
High Energy Transmission Grating Spectrometer (HETGS) for 75~ksec
during the decline of its 2002--2003 outburst.  The sensitivity of
this observation provides an unprecedented glimpse of a Galactic black
hole at about a tenth of the luminosity of the outburst peak.  The
continuum spectrum is well described by a model consisting of
multicolor disk blackbody ($kT \simeq 0.6$~keV) and power-law ($\Gamma
\simeq 2.5$) components.  X-ray reflection models yield
improved fits.  A strong, relativistic Fe~K$\alpha$ emission line is
revealed, indicating that the inner disk extends to the innermost
stable circular orbit.  The breadth of the line is sufficient to 
suggest that GX~339$-$4 may harbor a black hole with significant
angular momentum.  Absorption lines from H-like and He-like O,
and He-like Ne and Mg are detected, as well as lines which are likely
due to Ne~II and Ne~III.  The measured line properties make it
difficult to associate the absorption with the coronal phase of the
interstellar medium.  A scenario wherein the absorption lines are due
to an intrinsic AGN-like warm-absorber geometry --- perhaps produced by a
disk wind in an extended disk-dominated state --- may be more viable.
We compare our results to {\it Chandra} observations of the Galactic
black hole candidate XTE~J1650$-$500, and discuss our findings in
terms of prominent models for Galactic black hole accretion flows and
connections to supermassive black holes.
\end{abstract}

\section{Introduction}
The low mass X-ray binary (LMXB) GX~339$-$4 has been a very important
object in the study of Galactic black hole candidates (BHC), due to
several interesting properties.  It is a quasi-persistent/recurring
transient source, and therefore historically it has transited through
a wide range of luminosities and X-ray spectral states (Ilovaisky et
al.\ 1986, Grebenev et al.\ 1991, Miyamoto et al.\ 1991, Mendez \& van
der Klis 1997, Belloni et al.\ 1999).  The classification of
GX~339$-$4 as a BHC arose because its X-ray spectral and timing
behavior was very similar to that of dynamically determined BHCs
(e.g., Nova Muscae 1991, Miyamoto et al.\ 1993).  Recently, Hynes et
al.\ (2003) have used high resolution optical spectroscopy of Bowen
fluorescence lines from the secondary to arrive at a mass function of
${\rm M}_{BH} = 5.8\pm0.5~{\rm M}_{\odot}$ which clearly establishes
the black hole nature of the primary in GX~339$-$4.  The mass function
represents a lower limit to the black hole mass, but we will assume
this mass throughout as it is a conservative step.

Further interest in GX~339$-$4 is owed to the fact that the broad-band
spectrum, from radio through gamma-ray wavelengths, is likely
dominated by emission from the accretion flows around the central compact
object.  The secondary makes little contribution to the spectrum.  It
had been suggested, partly based upon possible fast time scale
X-ray/optical correlations (e.g. Motch et al.\ 1983), that even the
optical emission in the spectrally hard/high variability X-ray `low
state' was due to synchrotron emission from an extended outflow near
the compact object (Fabian et al.\ 1982, Imamura et al.\ 1990, Steiman
et al.\ 1990).  Evidence of a jet-like outflow in the low-luminosity
X-ray hard state was strengthened by the detection of radio emission
at $\approx 3-10$~mJy levels that is positively correlated with the
hard state X-ray flux (Fender et al.\ 1999, Wilms et al.\ 1999).  At
the opposite extreme of X-ray flux, the spectrally soft and moderately
variable `very high state' was first recognized as a state with
distinct properties in GX~339$-$4 (Miyamoto et al.\ 1991); later it
was also realized that outflows may occur in this state (Miyamoto et
al.\ 1995).

Fluorescent Fe~K$\alpha$ line emission can be a very powerful probe of
the accretion flow geometry in Galactic black holes.  Such lines are
very likely produced through irradiation of the inner accretion disk;
therefore, the breadth of such lines can serve as a trace of the inner
disk extent.  A primary prediction of advection-dominated accretion
flow (ADAF) models (see, e.g., Esin, McClintock, \& Narayan 1997) is
that the accretion disk should radially recede as the mass accretion
rate ($\dot{m}$) falls.  Following a long outburst of GX~339$-$4 in
2002--2003, we requested and were granted a Director's Discretionary
Time observation with {\it Chandra} in part to test this prediction as
the outburst decayed.  Although prior observations have found broad
lines suggesting that the disk remains close to the innermost stable
circular orbit (ISCO) in GX~339$-$4 both in rather bright phases and
deep in the low/hard state (see, e.g., Nowak et al.\ 2002), such
observations were limited by coarse spectral resolution and
sensitivity.  In contrast, the resolution of the {\it Chandra} High
Energy Transmission Grating Spectrometer (HETGS) allows broad
Fe~K$\alpha$ emission lines to clearly be revealed as intrinsically
broad (Miller et al.\ 2002a).

A second important motivation for our observation was to search for
evidence of a disk-driven outflow.  Preliminary indications for
absorption lines from low-Z elements in a BHC --- perhaps due to such
an outflow --- were first seen in two {\it Chandra}/HETGS observations
of the BHC XTE~J1650$-$500 (Miller et al.\ 2002b).  There is ample
observational evidence to suggest that such an outflow should exist in
Galactic black hole systems: a disk wind has been dramatically
revealed in the neutron star system Circinus X-1 (Brandt \& Schulz
2000), and disk-driven outflows may be common in AGN (Elvis 2000,
Morales \& Fabian 2002, Turner et al.\ 2003).  With gratings covering
the 0.5--10.0~keV band, the {\it Chandra}/HETGS is well suited to
searching for lines from low-Z elements and lines due to Fe L-shell
transitions.

\section{Observation and Data Reduction}
GX 339-4 was observed with {\it Chandra} for approximately 75~ksec,
starting on UT 2003 March 17.8.  The HETGS-dispersed spectrum was read
out with the ACIS-S array operating in timed exposure (TE) mode.  A
400-row subarray was used to reduce the frametime to $(400/1024)\times
3.2{\rm s} \simeq 1.25{\rm s}$.  As a result, the ``X'' pattern of the
dispersed spectra along the length of the ACIS array is truncated at
shorter wavelengths (higher energies) than if a full array had been
used.  The MEG spectra are truncated just below the O edge energy (the
O edge occurs 0.53~keV and is due to photoelectric absorption in the
inter-stellar medium, or ISM).  Similarly, the 400-row sub-array
limits the validity of the HEG spectra (which normally extending to
0.8~keV) to the 1.2--10.0~keV range.  Finally, a Y-coordinate
translation of $+$0.33 arcmin was used to move chip gaps that might
fall in the red wing of a broad Fe~K$\alpha$ emission line in the
first-order HEG spectra to the 9.5--10.5~keV range.

Apart from this observation of GX~339$-$4, the most sensitive {\it
  Chandra} spectra of an LMXB BHC were obtained through observations
of XTE~J1650$-$500.  XTE J1650$-$500 was observed with the {\it
Chandra}/HETGS on two occasions for 30~ksec each.  The first
observation started on UT 2001 October 5.9, and the second observation
started on UT 2001 October 29.0.  The HETGS-dispersed spectrum was
read-out with the ACIS-S array operating in continuous clocking (CC)
mode to avoid photon pile-up.  A ``gray'' filter (1 in 10 events were
telemetered in a region 100 columns wide) was put into effect in the
region of the zeroth order to also avoid telemetry saturation.  To
avoid any wear and tear on the nominal ACIS-S3 aimpoint while keeping
as much of the HEG Fe~K$\alpha$ line region on the
backside-illuminated ACIS-S3 chip, the nominal aimpoint was moved
4.0mm toward the top of the ACIS-S array, and a Y translation of
-1.33~arcmin was made.  These steps were likely overly conservative,
and had the negative consequence of placing chip gaps in the red wing
of any broad Fe~K$\alpha$ emission line.  Inspection of the data
reveals gain variations and/or CTI effects across some of the chips
that are not fully accounted for by the current calibration.

The {\it Chandra} datasets were reduced using tools in the CIAO version
2.3 analysis suite.  The highest level filtered event list (evt2) was
further refined by running the tool ``destreak'' on the ACIS-S4 chip
to remove detector effects.  Spectra were extracted at the nominal
instrumental resolution (HEG: 0.0025\AA, MEG: 0.005\AA) using the tool
``tgextract''.  Source and background spectra were extracted using the
nominal regions.  Standard redistribution matrix files (rmfs) from the
CALDB database were used to generate ancillary response files (arfs)
and for spectral fitting; arf files were made using the script
``fullgarf''.  Lightcurves of the dispersed counts were made
using the tool ``lightcurve''.

Only background-subtracted spectra were considered for analysis;
nominal background regions were selected during the source spectrum
extraction.  The high-resolution spectra were analyzed using
ISIS version 1.0.50 (Houck \& Denicola 2000).  Analysis of the broad-band
spectrum and Fe~K$\alpha$ emission line was made using XSPEC version
11.2 (Arnaud \& Dorman 2000).  After inspecting the first-order spectra
individually, the background-subtracted first-order MEG spectra and
arfs were added using the tool ``add\_grating\_spectra'' prior to
analysis; the same procedure was performed on the
background-subtracted first-order HEG spectra and arfs.

\section{Analysis and Results}

The lightcurve of the dispersed spectrum of GX 339$-$4 is remarkably
constant (see Figure 1), and so we consider the time-averaged spectra
for analysis.  The MEG spectra are more sensitive below 2~keV, however
deviations are clear below 1~keV which are likely due to the
well-documented carbon build-up on the ACIS detectors.  At the time of
writing, there is no way to correct for this effect in grating
spectra.  Examining the individual ISM photoelectric absorption edges
in the source spectrum with local models consisting only of power-law
continua with simple edge functions, we find that the neutral (atomic)
O, Fe L3 and L2, Ne, Mg, and Si edges indicate abundances of 0.75--1.0
(relative to Solar) along this line of sight.  We regard these
measurements with caution, however.  Not only is the effective area
below 1~keV affected by the carbon build-up (for a discussion, see
Marshall et al.\ 2003), but the Si range is distorted by an ``inverse
edge'' near 2.07~keV, likely due to the Ir coating on the mirrors.
Due to these calibration uncertainties, we have characterized the
broad-band continuum properties using the HEG spectra as their broader
energy range should limit the relative importance of these systematic
effects.  As the calibration uncertainties below 1~keV are rather
broad-band in nature, local power-law models were used to characterize
the continuum when measuring the properties of narrow absorption
lines.  All line measurements were made using the MEG spectra, due to
the higher effective area of the MEG in this range.

\subsection{The Continuum Spectrum and Relativistic Fe~K$\alpha$
  Emission Line}

We made fits to the spectra from GX~339$-$4 and XTE J1650$-$500 in the
1.2--10.0~keV range (see Section 2).  In all fits, an inverse edge was
included at 2.07~keV with $\tau = -0.13$ to account for a calibration
error due to the Ir coating on the mirrors (see, e.g.,
http://asc.harvard.edu/cal/Links/Acis/acis/Presentations/
head\_nm\_meet/section2.html).  This value for $\tau$ is largely
model-independent and within the range commonly measured.

The results of all broad-band spectral fits are listed in Table 1.  A
standard model for the continuum spectra of BHCs consists of
multi-color disk (MCD) blackbody and power-law components, modified by
photoelectric absorption in the ISM (fits reported in this paper used
the ``phabs'' model within XSPEC).  The MCD model is based on the
Shakura \& Sunyaev (1973) accretion disk model (Mitsuda et al.\ 1984),
while the power-law component is a phenomenological description of
Comptonization and/or synchrotron emission.  We first fit the spectra
considered in this work with the MCD plus power-law model, with the
power-law index constrained to be within $\Delta(\Gamma) \leq 0.2$
of the value measured via fits to simultaneous {\it RXTE} spectra in
the 3--200~keV band.  Joint fits to the simultaneous {\it Chandra} and
{\it RXTE} spectra will be presented in separate work.  Unabsorbed
fluxes were calculated on the standard 0.5--10.0~keV band to best
correspond to other {\it Chandra} and {\it XMM-Newton} LMXB and BHC
observations made using more common instrumental offsets and
parameters.  The spectra of XTE J1650$-$500 are affected by chip gaps
below the Fe~K$\alpha$ emission line range due to the Y-offset chosen
for these observations, and so we only characterize the spectra of
XTE~J1650$-$500 with this phenomenological but standard model.  In
contrast, the spectrum of GX~339$-$4 is free of such complications;
strong deviations consistent with a broad Fe~K$\alpha$ emission line
are revealed in the data/model ratio (see Figure 2).

We therefore included the ``Laor'' line model to the spectral model
for GX~339$-$4.  This model describes the line profile expected when
an accretion disk orbiting a black hole with non-zero angular momentum
($a = j = cJ/GM^{2}$) is irradiated by a source of hard X-rays.  The
line energy, inner disk emissivity index $q$ ($J(r) \propto r^{-q}$
where $J(r)$ is the emissivity and $q=3$ is expected for a standard
thin disk), the inner and outer radii of the line emitting region, the
inner disk inclination, and the line normalization are free
parameters.  The outer emission radius was fixed to $R_{out} =
400~R_{g}$ in fits with this model ($R_{g} = GM/c^{2}$); all other
parameters were left as free parameters.  The addition of this
component significantly reduces the $\chi^{2}$ fit statistic; the
F-statistic for the inclusion of this model ($3.6\times 10^{-10}$)
indicates that the Laor model is required at the $6.3\sigma$ level of
confidence.  The addition of a smeared edge (``smedge'' within XSPEC)
is not statistically required, likely due to the low effective area of
the HEG above 8~keV.

It is clear that the spectrum of XTE J1650$-$500 evolved significantly
between the observations.  In both spectra, moderate disk color
temperatures are measured ($kT = 0.610(4)$~keV and $kT =
0.570(1)$~keV, respectively).  In the first observation a moderately
hard power-law ($\Gamma = 2.5^{+0.1}_{-0.2}$) comprises 21\% of the
unabsorbed flux in the 0.5--10.0~keV band; however, in the second
observation the power-law is very steep ($\Gamma = 5.4$) and
comprises $<3$\% of the total flux.  Indeed, the total flux
dropped by approximately 40\% between the first and second
observations.  The spectral parameters measured during the first
observation of XTE~J1650$-$500 indicate that the source was likely
observed in the ``very high'' or ``intermediate state'', which may
only be manifestations of the same state at different source fluxes
(see, e.g., Homan et al.\ 2001).  The parameters measured in
the second observation, however, indicate that XTE J1650$-$500 was
likely observed in the ``high/soft'' state.

The spectral parameters measured from GX~339$-$4 are more like those
measured in the first observation of XTE J1650$-$500 (the power-law is
13\% of the total 0.5--10.0~keV flux in the spectrum GX 339$-$4).
However, GX~339$-$4 was observed at a flux 7.4 times and 4.5 times
lower than measured in the first and second observations of
XTE~J1650$-$500.  Indeed, the observation of GX~339$-$4 was made
following an extended spectrally soft state as the source transitioned
to the ``low/hard'' state.

The {\it Chandra}/HETGS provides dispersive spectroscopy through the
Fe~K$\alpha$ line range, making it possible to learn whether an
apparently broad line is intrinsically broad, comprised of a few
narrow lines, or perhaps a combination of an intrinsically broad line
with narrow components superimposed.  This ability was exploited to
show that the broad Fe~K$\alpha$ emission line in Cygnus X-1 does
indeed contain an intrinsically broad component likely shaped by
relativistic effects (Miller et al.\ 2002a).  Unlike Cygnus X-1, we
find no narrow lines amidst the broad line profile in GX~339$-$4;
however the HETGS spectrum again demonstrates that the broad
Fe~K$\alpha$ line is intrinsically broad.  The Laor line model
indicates that the inner disk edge may extend as close as $R_{in} =
2.5_{-0.3}^{+2.0}~R_{g}$ to the black hole, which would imply black
hole spin as $R_{ISCO} = 6~R_{g}$ for $a=0$ black holes (while
$R_{ISCO} = 1.24~R_{g}$ for $a=0.998$).  A moderate inner disk
inclination is measured, consistent with prior X-ray and optical
studies of GX~339$-$4.  The line equivalent width is high
(W$_{K_{\alpha}} = 600^{+200}_{-100}$~eV); this is broadly consistent
with predictions from reflection models if the disk is highly ionized
(Ross, Fabian, Young 1999).  The measured line centroid energy of
E$=6.8^{+0.2}_{-0.1}$~keV is consistent with a combination of He-like
Fe~XXV and H-like Fe~XXVI.

The normalization of the MCD model includes the inner radius of the
accretion disk explicitly ($N = ((R_{in}/{\rm km})/(d/10{\rm
kpc}))^{2}~{\rm cos}(i)$), and is therefore often used to infer the
inner disk radius in BHCs.  Assuming a distance of 4--5~kpc and an
inclination of $i=45^{\circ}$, our best-fit normalization gives
$R_{in} = 22-27$~km.  Assuming a black hole mass of
$M_{BH} = 5.8~M_{\odot}$ as per Hynes et al.\ (2003) for the black
hole mass, this corresponds to $R_{in} = 2.6-3.2~R_{g}$, consistent
with measurements using the Laor model.  Although a recent study of 49
spectra spanning roughly one month of the 2002 outburst of the BHC
4U~1543$-$475 reveals general agreement between MCD-derived and
Laor-derived estimates for the inner disk radius (Park et al.\ 2003),
we regard the radii derived from the Laor model as more robust for two
reasons.  First, the MCD model is only an approximation to the Shakura
\& Sunyaev (1973) disk model in that it neglects the inner boundary
term (see, e.g., Titarchuk \& Shrader 2002).  Second, the inner disk
temperature and radius measured using the MCD model are subject to
Comptonization and angle-dependent opacity distortions (Shimura \&
Takahara 1995; Merloni, Fabian, \& Ross 2000).

We also made fits with the constant density ionized disk (CDID) and
``pexriv'' reflection models (see Table 1), which are more physical
models describing the illumination of the accretion disk by a
power-law source of hard X-rays (CDID: Ross, Fabian, \& Young 1999;
pexriv: Magdziarz \& Zdziarski 1995) .  The CDID model is particularly
well-suited to high ionization regimes, and so we used the results of
fits made with the CDID model to guide our fits with pexriv (pexriv
does not Comptonize absorption edges and is therefore more
approximate).  An important difference between these reflection models
is that the CDID model explicitly includes Fe~K$\alpha$ line emission,
while a separate line component must be added to the total model to
make proper fits with pexriv.  In both cases, we convolved (or,
``smeared'') the reflection model with the Laor line element to
account for the strong Doppler and gravitational shifts expected near
to the black hole.  The reflected spectra were fit in addition to an
MCD component, as a disk component is not included explicitly in
either model.  The reflected spectra were smeared with the Laor
element for two reasons.  First, prior fits to the Fe~K$\alpha$ line
with the Laor model (but with a less physical continuum model) already
suggest that GX 339$-$4 may harbor a black hole with significant
angular momentum.  Second, the difference between the Laor line
profile for $R_{in} > 6~R_{g}$ and a Schwarzschild line profile (e.g.,
Fabian et al.\ 1989) is sufficiently small that fits yielding
$R_{in,~Laor} > 6~R_{g}$ must serve as a potential sign of a very low
spin parameter.

Fits with the CDID model yielded a significant improvement over fits
with the phenomenological model consisting of MCD, power-law, and Laor
components (see Table 1).  The best-fit ionization parameter is
well-constrained and indicates a highly-ionized disk reflector: ${\rm
log}(\xi) = 4.0\pm 0.4$ (where $\xi = L_{X}/nr^{2}$ and $n$ is the
hydrogen number density).  The reflection fraction is measured to be
$R=1.2^{+2.0}_{-0.5}$, indicating that the disk intercepts roughly
half of the hard X-ray flux ($R = \Omega/2pi$).  The ionization
parameter measured is in agreement with the iron ionization state
measured via the Laor model.  The best-fit Laor smearing parameters
($R_{in} = 1.3^{+1.7}_{-0.1}~R_{g}$, $q=3.0^{+0.7}_{-0.4}$,
$i=30^{\circ}\pm 15^{\circ}$) again suggest a disk which extends very
close to the black hole, suggesting the black hole may have a high
spin parameter.  In contrast to the more phenomenological model,
however, the smeared CDID fits do not require a high disk emissivity
index.

The results of fits made with the CDID model were used to guide fits
with pexriv.  We fixed the disk temperature to $T = 1.2\times
10^{7}$~K (1~keV) to correspond with the expected temperature of an
ionized accretion disk skin (Young et al.\ 2001), and ${\rm log}(\xi)
= 4.0$ to the above CDID result.  We fit pexriv as a ``pure''
reflection spectrum (e.g., without including the power-law continuum);
the irradiating power-law index and normalization were fixed to
correspond to the separate power-law component.  The inclination in
pexriv was fixed to correspond to the same parameter in the
relativistic smearing function.  Finally, we fixed the high energy
cut-off of the power law to E$=$200~keV in view of the lack of an
obvious cut-off in the 3--200~keV {\it RXTE} spectrum.  As with the
CDID model, fits with pexriv are statistically preferred to fits with
the initial phenomenological model (see Table 1 and Figures 3 and 4).
The reflection fraction is measured to be $R = 0.90\pm 0.4$, again
indicating that the disk intercepts roughly half of the incident flux.
The relativistic smearing parameter values were set to be equivalent
to the parameters in the independent Laor line model.  Fits with the
Laor line once again indicate an ionized line (E$=6.97_{-0.05}$~keV)
and a small inner disk radius $R_{in} = 2.0^{+2.2}_{-0.7} R_{g}$.

\subsection{The High Resolution Spectrum of GX~339$-$4}

We analyzed the combined first-order MEG spectrum of GX~339$-$4 in
narrow 2\AA~ slices.  The continuum in each slice was fit with a
simple power-law model, modified by an edge due to photoelectric
absorption by neutral (atomic) elements in the ISM where appropriate.
Line features were fit with Gaussian models.  Line identifications are
based on the calculations and line lists tabulated by Verner, Verner,
\& Ferland (1996) and by Behar \& Netzer (2002).  Table 2 lists the
line identifications and parameters for absorption lines detected in
the spectrum of GX~339$-$4 (as well as XTE~J1650$-$500, against which
we later compare the spectrum of GX~339$-$4), as well as upper-limits
for lines not clearly detected.   The column density estimates were
derived using the relation:\\

\begin{center}
$W_{\lambda} = \frac{\pi e^{2}}{m_{e} c^2} N_{j} \lambda^{2} f_{ij} =
8.85 \times 10^{-13} N_{j} \lambda^{2} f_{ij}$\\
\end{center}

\noindent where $N_{j}$ is the column density of a given species,
$f_{ij}$ is the oscillator strength, $W_{\lambda}$ is the equivalent
width of the line in cm units, and $\lambda$ is the wavelength of the
line in cm units (Spitzer 1978).  We have taken all $f_{ij}$ values
for all resonant lines from Verner, Verner, \& Ferland (1996) and
Behar \& Netzer (2002).  In using this relation, we are assuming that
the lines are optically thin and that we are on the linear part of the
curve of growth.  Fits to the spectrum of GX~339$-$4 are shown in
Figures 5 and 6.

A number of relatively strong X-ray absorption lines are clearly
detected in the high resolution spectrum.  We identified the observed
lines using a boot-strap approach. One of the strongest lines is
found at 13.442(1)\AA~, and we identify this line as the
Ne~IX He-$\alpha$ resonance line expected at 13.447\AA.  Absorption
lines observed at $11.541^{+0.006}_{-0.004}$\AA~ and 10.998(3)\AA~ are
then very likely to be the Ne~IX He-$\beta$ and He-$\gamma$ resonance
lines expected at 11.547\AA~ and 11.000\AA, respectively.  

With the clear identification of the He-like Ne~IX resonance series,
we then searched for He-like and H-like resonance absorption lines
from Ne and other low-Z elements.  Lines detected at 18.964(1)\AA~ and
16.003(4)\AA~ correspond with the O~VIII Ly-$\alpha$ and Ly-$\beta$
lines expected at 18.967\AA~ and 16.006\AA~,
respectively.  The line observed at 18.623(1)\AA~ corresponds to the
O~VII He-$\beta$ line expected at 18.627\AA~ (the He-$\alpha$ line is
expected at 21.602\AA, where the sensitivity of the spectrum is low).
The line detected at $9.166^{+0.001}_{-0.002}$\AA~ corresponds to the
Mg~XI He-$\alpha$ line expected at 9.169\AA.  The strongest of these
highly ionized lines have velocity widths $\leq 400$~km/s, and
apparent blue-shifts of approximately 100~km/s.

The Si He-like resonance line series is not clearly detected.  With
the exception of O, lines from H-like series are not detected.  A
number of lines from the Fe L shell (especially Fe~XVI and Fe~XVII)
are expected to be especially strong in the 12--17\AA~ range, but
these lines are also absent in the spectrum.  We measured upper limits
for H-like lines and particular Fe~XVI and Fe~XVII lines; these
upper-limits are also reported in Table 2.

Two especially strong lines are detected at 14.606(1)\AA~ and
14.504(2)\AA.  These lines are not plausibly associated with any
He-like or H-like lines from low-Z elements.  The O VIII 1s-6p and
1s-7p transitions are expected at 14.634\AA~ and 14.524\AA~
(respectively), but this would imply larger blue-shifts (400-600~km/s)
for these lines than for the detected O VIII 1s-2p and O VIII 1s-3p
lines ($<100$~km/s).  Moreover, the 1s-4p and 1s-5p lines are not
detected, which makes any identification with O VIII very problematic.
Similarly, these lines cannot plausibly be associated with Fe L-shell
transitions occurring in this range, as the strongest Fe L-shell lines
expected for a wide range in plasma temperature, density, and
ionization parameter (see, e.g., Kallman \& McCray 1982) are not
detected and the upper limits are strict.  In particular, although
these lines are close to Fe~XVIII transitions, other Fe~XVIII
transitions which should be stronger are not detected.

Behar and Netzer (2002) have recently calculated the $1s-2p$ resonance
line wavelengths and oscillator strengths expected for abundant
elements up to Fe.  The Ne~II resonance line occurs at 14.631\AA~ and
the Ne~III line occurs at 14.526\AA.  The wavelengths calculated by
Behar and Netzer (2002), if accurate, would imply blue-shifts of
400--500~km/s.  If the lines at 14.631\AA~ and 14.526\AA~ are Ne~II and
Ne~III resonance absorption lines, their velocity widths (500~km/s and
400~km/s, respectively) imply that the absorption may not originate in
the ISM.  Thermal broadening in the cool phase of the ISM is expected
to be lower than 10~km/s (e.g., Vidal-Madjar et al.\ 1982).  Such
narrow lines would be highly saturated, and column densities much
larger than those in Table 2 would be required to account for the
measured equivalent widths.  Lines similar to those we identify as
Ne~II and Ne~III may have have detected in other sources, though not
as clearly.  A feature which may be a blend of the lines we have
observed in GX 339$-$4 was found in a {\it Chandra}/LETGS spectrum of
Cygnus X-2 and ascribed to Ne II (Takei et al.\ 2002).  In an analysis
of a {\it Chandra}/LETGS spectrum of 4U 0614$+$091 (Paerels et al.\
2001), a possible absorption line was found at 14.45\AA.

\subsection{The High Resolution Spectra of XTE J1650$-$500}

The two sensitive spectra of XTE J1650$-$500 obtained with the {\it
Chandra}/HETGS provide an important background against which to
contrast the results of our analysis of the GX~339$-$4 spectrum.
Therefore, we analyzed these spectra in exactly the same manner used
to analyze the spectrum of GX~339$-$4.  Results from a preliminary
analysis of the spectra of XTE J1650$-$500 were previously reported by
Miller et al.\ (2002b); however, the results we report herein are based
on a more refined reduction and analysis.  The parameters for both
detected lines and upper limits are reported in Table 2, and fits to
the spectra of XTE J1650$-$500 are shown in Figure 7.

The Ne~IX He-$\alpha$ line is not detected in the first spectrum of
XTE~J1650$-$500, and the upper limit is strict.  A weak line is
detected at about the $4\sigma$ level of confidence in the second
spectrum.  The Mg~XI He-$\alpha$ line is not detected in the spectrum
from either observation, and once again the upper-limits are constraining.

The line we have identified as Ne~II is detected in both spectra of
XTE J1650$-$500, at 14.606(3)\AA~ and 14.609(6)\AA~ in the first and
second observations, respectively.  The line equivalent width
(and, therefore, the implied column density) is measured to decrease
slightly between the first and second observations.  However, the
Ne~III line expected at 14.526\AA~ and found at 14.504(2)\AA~ in the
spectrum of GX~339$-$4 is not clearly detected in either spectrum of
XTE~J1650$-$500.  It is worth noting that the line herein identified
as Ne~II was originally attributed to an Fe~XVIII transition in Miller
et al.\ (2002b); for the reasons given in the previous section the
prior identification is likely in error.

\section{Discussion}
\subsection{The Fe~K$\alpha$ Emission Line and Black Hole Spin}

Our fits to the Fe~K$\alpha$ emission line profile, using the Laor
line model (in conjunction with either a simple MCD plus power law
model, or with smeared, ionized reflection models), clearly indicate a
relativistically broadened line.  The line and smearing parameters
indicate emission peaked towards small radii, and an inner emission
radius $R_{in} < 6~R_{g}$, which may serve to indicate that GX 339$-$4
harbors a black hole with significant spin.  Strong evidence for black
hole spin in BHCs based on skewed Fe~K$\alpha$ line emission has been
reported in XTE J1650$-$500 (Miller et al.\ 2003) and GRS~1915$+$105
(Martocchia et al.\ 2003).  

The evidence for spin in GX~339$-$4 is not as strong as that indicated
by the Fe~K$\alpha$ emission line found in the spectrum of
XTE~J1650$-$500 with {\it XMM-Newton} (Miller et al.\ 2003).  Fits to
the spectrum of XTE~J1650$-$500 including a Laor line model and the
pexriv reflection model ruled-out $R_{in} = 6~R_{g}$ at more than the
$6\sigma$ level of confidence --- strongly suggesting a black hole
with a high spin parameter.  Indeed, the line profile found in this
spectrum of GX~339$-$4 is only clearly visible above the continuum at
energies above approximately 5~keV; in contrast, the line revealed in
the {\it XMM-Newton} spectrum of XTE~J1650$-$500 extends down to
approximately 4~keV.  The normalization of the MCD model may also
support the finding of a fairly small inner disk radius for
GX~339$-$4.  Although it can be problematic to directly translate a
fitted disk radius to an exact physical radius, the normalization of
the MCD model gives an inner disk radius of $R_{in} = 2.6-3.2 R_g$ for
a $5.8~M_\odot$ black hole at 4--5~kpc.  Thus, although other fits to
this spectrum and its prominent Fe~K$\alpha$ emission line suggest
that GX~339$-$4 harbors a black hole with a high spin parameter, the
data must be regarded as indicative but not definitive.

It is interesting to note that most of the systems for which some
degree of black hole spin is implied --- either through Fe~K$\alpha$
emission line profiles or high frequency QPOs --- have also been
observed to eject highly relativistic jets (e.g., GRO~J1655$-$40, XTE
J1550$-$564, GRS~1915$+$105, Fabian \& Miller 2002).  For a distance
of 4--5~kpc, the extended radio emission previously observed in
GX~339$-$4 only gives a weak constraint on the jet velocity ($v/c \geq
0.1$) because the source was observed infrequently (Fender et al.\
1997).  Future observations of GX~339$-$4, if made with sufficient
frequency and at high angular resolution, may be able to measure more
accurate jet ejection velocities.

Finally, it has recently been suggested that broad, red-shifted lines
may be the product of Thomson scattering in an optically-thick,
wide-angle or quasi-spherical outflow with $v/c \sim 0.3$ (Titarchuk,
Kazanas, \& Becker 2003).  There are many problems with such a model.
The hot inner disk component and high frequency QPOs (${\rm few}
\times 100$~Hz, likely tied to frequencies in the inner disk) seen in
BHCs at the highest implied accretion rates (for a review, see
McClintock \& Remillard 2003) should not be visible through an
optically-thick outflow.  Moreover, considering only the Eddington
limit equation, the mass outflow rate and the requirement that the
outflow be optically-thick, and realizing that line emission
unaffected by the relativistic regime near the black hole must occur
at radii greater than approximately $100~R_{g}$, this model requires
the mass outflow rate to exceed the mass inflow rate by at least a
factor of a few.  In view of these difficulties, we conclude that
relativistic broadening remains the most viable explanation for broad
Fe~K$\alpha$ emission lines like that revealed in GX~339$-$4.

\subsection{The Nature of the Inner Accretion Flow Geometry}

The observation of GX~339$-$4 discussed here is likely best described
as having caught the source in an 'intermediate state', as the source
transitioned to the low/hard state.  There is significant hard X-ray
emission, but the spectrum also reveals a more substantial disk
component than typical hard states.  A number models suggest that
falling, hardening flux may be the signature of a disk that is
receding radially and being replaced by a hot, Comptonizing region in
the central accretion region.  Other models predict that these
signatures may be the signature of jet-dominated state.  

It is clear that the inner disk is not recessed in this low-luminosity
intermediate state.  Assuming an $M = 5.8~M_{\odot}$ black hole and a
distance of $d=4-5$~kpc, we observed GX 339$-$4 at $\dot{m}_{Edd}
\simeq 0.1-0.2$.  The model for Galactic black hole state transitions
described by Esin, McClintock, \& Narayan (1997) predicts that the
inner disk should be mildly recessed at similar accretion rates and/or
in the intermediate state.  At accretion rates lower than $\dot
m_{critical} = 0.08$, this model predicts that the inner disk edge
should radially recede to $R > 100 R_g$ and be replaced by an inner
ADAF.  The spectrum of GX~339$-$4 discussed herein shows that
ADAF-based models for BHCs must be adapted to accommodate --- at a
fairly low fractional Eddington luminosity --- an intermediate state
that has a disk extending to the ISCO and a reasonably hard spectrum
that is rather comparable to "pure" hard (putatively ADAF) states. Any
significant change of the radius of the inner edge of the disk from
the small values suggested by these observations to the large values
posited by the ADAF models must occur at luminosities lower than those
observed here.  Some observations of BHCs suggest that the inner disk
may recede very gradually (e.g., Orosz et al.\ 1995; Tomsick, Kalemci,
\& Kaaret 2003).  However, our observations reasonably might be
considered to be more consistent with models wherein changes of the
inner disk radii are minimal, such as the transition to the hard state
being driven by the growth of a corona and subsequent ionization of
the outer layers of the disk atmosphere (Young et al.\ 2001).

Although our observations likely occurred before the full transition to
a radio loud, X-ray hard state, radio observations of GX~339$-$4 made
7 days prior to our {\it Chandra} observation detect optically-thin
(likely jet-based synchrotron) emission at a flux density of 0.2 mJy
at 4.8~GHz (Gallo et al.\ 2003).  In XTE J1118$+$480 and other sources
(including GX~339$-$4 at flux levels comparable to the flux measured
in this observation), hard X-ray emission in the low luminosity states
has been modeled as due to synchrotron emission from the jet (e.g.,
Markoff, Falcke, \& Fender 2001, Markoff et al.\ 2003).  If hard X-ray
emission in BHCs is due solely to jets, very low reflection fractions
should be observed.  Lorentz factors of $\gamma=2-3$ are required to
make the the hard X-ray emission in jets, creating a flux enhancement
of $\gamma^{3}$ within the jet cone due to beaming.  Thus, 8--27 times
less flux is expected to irradiate the disk than if the hard X-rays
are generated by an isotropic source.  The high reflection fractions
we have measured in this spectrum of GX~339$-$4 ($R \sim 1$, see Table
1) are inconsistent with a scenario in which the hard X-ray emission is
beamed away from the disk.  Although jet models are applied across a
large range in Eddington luminosity, from $L_{X}/L_{Edd.} \leq
10^{-10}$ in Sgr A* (Baganoff et al. 2003) to $L_{X}/L_{Edd.} \simeq
10^{-2}-10^{-1}$ as per the low/hard state in BHCs, it is not
clear that jet models are appropriate in this state, as the disk flux
is still a sizable fraction of the total.  Regardless of whether or
not jet models are appropriate in the intermediate state, this
observation of GX~339$-$4 requires that a separate, non-jet hard X-ray
component (very likely, a Comptonizing corona) generate the majority of
hard X-ray emission.

\subsection{Evidence for an Intrinsic Warm Absorber}

In an optically-thin plasma with a temperature near $T \simeq
10^{5-6}$~K, the He-like ions of O, Ne, and Mg are expected to be
important (e.g., Mazzotta et al.\ 1998).  Some simple calculations suggest
that the lines we have observed could plausibly be explained through
absorption in the coronal phase of the ISM, which is at a temperature
near $10^{6}$~K.  If we assume 1) the density of coronal gas is
$n_{coronal} = 1.0 \times 10^{-2}~{\rm cm}^{-3}$, 2) coronal gas
occupies 50\% of the ISM ($f_{coronal} = 0.5$), 3) the abundance of Ne
relative to H is $A_{Ne} = 1.0 \times 10^{-4}$, and 5) the distance to
GX~339$-$4 is $d=4.0$~kpc, then $N_{Ne} \simeq n_{coronal} \times
f_{coronal} \times A_{Ne} \times d \simeq 6 \times 10^{15}~{\rm
cm}^{-2}$.  This value is rather close to that reported for GX~339$-$4
in Table 2.  Moreover, a solar abundance plasma in collisional
ionization equilibrium (Mazzotta et al.\ 1998) at T$\sim 2-3
\times 10^6$ K comes reasonably close to matching the relative column
densities of the observed ions while not violating the upper limits.
Although these factors point to absorption in the ISM as a viable
explanation for the observed absorption spectrum in GX 339$-$4
(indeed, additional deep observations of Galactic black holes are
required to definitively rule-out this possibility), there are a
factors which are plainly inconsistent with this explanation.
Below, we discuss some difficulties with ascribing the absorption
lines to the coronal ISM, and suggest that a warm absorbing geometry
near to the black hole system likely provides a better explanation.

First, on examination of the He-like Ne series in the GX 339$-$4
spectra, the equivalent widths follow the expected ratios based on
oscillator strength.  This indicates that the Ne absorption lines are
not saturated and retain a true shape.  The width of the resolved Ne
He-$\alpha$ line is $350^{+80}_{-90}$~km/s (FWHM), consistent with the
Mg He-$\alpha$ and O Ly-$\alpha$ lines (the O line is likely partially
saturated).  If the lines were intrinsically narrower than 200~km/s,
they would fall on the flat part of the curve of growth, giving
smaller ratios between the He-$\alpha$ and He-$\beta$ lines and
requiring much higher column densities.  The Ne~II and Ne~III lines
are slightly broader than even the Ne IX He-$\alpha$ line.  Thermal
broadening in the coronal phase of the ISM is expected to be
approximately 100~km/s, which is significantly below the widths we
measure from the strongest absorption lines.  A local, intrinsic warm absorber
with moderate turbulent velocity broadening may be a better
explanation for the width of the lines observed.

Second, the strongest highly ionized lines in the spectrum of GX
339$-$4 have blue-shifts of approximately 100~km/s.  While the
absolute calibration uncertainty in the HETGS wavelength grid (0.05\%)
is comparable to the shifts observed, the HETGS calibration web page
(http://space.mit.edu/HETG/flight.html) indicates that this
uncertainly is likely overestimated and that shifts observed in many
lines over a range in wavelength are likely reliable.  If an intrinsic
warm absorber in GX~339$-$4 is created through an accretion disk wind,
then a small flow velocity into our line of sight would be expected,
qualitatively consistent with the data.  It is important to note that
the observed shifts are not due to systemic velocities: Hynes et al.\
(2003) find that GX~339$-$4 is likely moving away at 30~km/s
(Sanchez-Fernandez et al.\ 2002 find that XTE J1650$-$500 is moving
away at $19\pm 3$~km/s).

Third, if the absorption lines in GX 339$-$4 are due to the coronal
ISM, it is hard to reconcile the clear detection of many ionized X-ray
absorption lines in GX~339$-$4 with the non-detection of many of the
same features in XTE J1650$-$500.  The two sources are similar in some
very important ways.  Both GX~339$-$4 and XTE~J1650$-$500 lie within
the Galactic plane: $b_{GX~339-4} = -4.3$ and $b_{1650} = -3.4$.  The
equivalent neutral hydrogen density along the line of sight to these
sources is expected to be very similar: $N_{H, GX~339-4} = 5.3 \times
10^{21}~{\rm cm}^{-2}$ and $N_{H, 1650} = 4.9 \times 10^{21}~{\rm
cm}^{-2}$ (based on H~I measurements in radio by Dickey \& Lockman
1990).  The measured values are very similar to expectations.  Indeed,
the column to GX 339$-$4 is measured to be {\it lower} than that to
XTE~J1650$-$500 (see Table 1; the small discrepancies between the H~I
and X-ray measurements may arise due to a number of factors,
including, e.g., the coarse angular resolution of single-dish H~I
surveys, neglecting gas in molecular form, and uncertainties in the
ACIS calibration due to the build-up of carbon-based contaminant).  We
therefore expect that the ISM should be rather similar along the line
of sight to each source.  Prior observations in optical and X-ray
bands suggest that both systems are seen at relatively low ($i \leq
45^{\circ}$) binary and inner disk inclinations (GX 339$-$4: Hynes et
al. 2003, Nowak et al. 2002; XTE J1650-500: Miller et al. 2002c,
Sanchez-Fernandez et al. 2002).  Finally, the distances to these
sources are likely similar.  The {\it RXTE}/ASM measured a peak
count-rate of 70~c/s (1.5--12~keV) during the 2002--2003 outburst of
GX 339-4, and a peak count rate of 40~c/s during the 2001--2002
outburst of XTE J1650$-$500.  Assuming that similar fractional
Eddington luminosities were achieved and similar spectral shapes at
peak, then the distance to these sources is not likely to differ by
more than $\sqrt(70/40) \simeq 1.3$.

These similarities make it hard to explain how the column density of
Ne~IX is measured to be 30 times higher in GX 339$-$4 than in the
first observation of XTE J1650$-$500, if the lines are due to
absorption in the coronal ISM.  The source distances certainly do not
differ by a factor of 30.  It is also very unlikely that the distance
and abundance of Ne~IX in the ISM both differ by $\sqrt(30)$.
Moreover, the column density of Ne~IX is measured to increase
significantly between the first observation (made at a flux of
0.4~Crab) and the second observation (made at a flux of 0.25~Crab) of
XTE J1650$-$500 (the column density in the second observation is 2.3
times higher than the upper-limit in the first observation).  Note
that the observation of GX 339$-$4 -- wherein Ne~IX is clearly
detected -- occurred at a flux below 0.1~Crab.  

In view of these difficulties, a warm absorber (see, e.g., Reynolds
1997) intrinsic to GX~339$-$4 that is created by a disk wind or a shell
ejection event seems to provide a much better explanation for the
discrepant and changing line fluxes than absorption within the ISM.
Alternatively, the disk may not contribute a substantial outflow, but
the ionization of an absorber may fall as the central source flux
decays, allowing lines to be observed as the outburst decays.  Thus,
we are left with two possibilities: either the line of sight to GX
339-4 is special, for instance passing through a supernova remnant, or
else the absorbing gas is photoionized circumstellar material.  At the
time of writing, we have found no published claims for a supernova
remnant along this line of sight.  There is evidence for an outflowing
absorber in XTE J1650$-$500 as well; however, we have chosen to focus
this discussion and modeling efforts on GX~339$-$4 as the evidence is
much stronger.

\subsection{Photoionized Plasma Models}

X-ray absorption lines have been detected in Galactic black holes with
low-mass donors observed with {\it ASCA} (GRO~J1655$-$40: Ueda et al.\
1998, GRS 1915$+$105: Kotani et al.\ 2000).  In contrast to GX 339$-$4
and XTE J1650$-$500, these sources are viewed at high inclinations,
and the observed lines are consistent with Fe~XXV and Fe~XXVI.  The
absorbing geometry in these sources is very likely the hot
Comptonizing corona thought to contribute to hard power-law emission,
or an ionized skin above the accretion disk.  If the lines in the
relatively face-on GX~339$-$4 and XTE J1650$-$500 are due to intrinsic
warm absorbers, the absorber is likely of a different nature.
Certainly, the fact that only lines from low-Z elements are observed
signals an absorber which is likely 10--100 times colder than a
standard corona.

The warm absorber required might be analogous to those seen in some
Seyfert galaxies, perhaps fed through a disk-driven outflow (Elvis
2000, Morales \& Fabian 2002, Turner et al.\ 2003).  In contrast to a
recent model for AGN spectra which relies on an optically-thick
outflow (King \& Pounds 2003); it is clear that any disk-driven
outflow in GX~339$-$4 and XTE J1650$-$500 must be {\it optically
thin}.  First, the spectra are well-described by an optically thin gas
in photoionization equilibrium.  Second, the inner disk is revealed
through the hot disk component and relativistic Fe~K$\alpha$ line, both
of which would be obscured by an optically thick outflow.

To further investigate this idea we have computed the equivalent
widths of the observed absorption lines using the atomic physics
packages described in Raymond (1993).  Briefly, these include heating
by photoionization and Compton scattering and cooling by line and
continuum emission and Compton scattering.  The predicted ionization
state is combined with a curve of growth (Spitzer 1978) for a chosen
velocity width to determine the equivalent widths.

We specify the ionizing spectrum using the blackbody and power-law
parameters in Table 1.  Placing a $10^{7}$~cm thick slab with a
density of $3 \times 10^{13}~\rm cm^{-3}$ at a distance of $2 \times
10^{11}$~cm from the X-ray source gives an ionization parameter $\xi =
L/n r^2$ of 70.  It predicts about the right Ne IX equivalent widths
for a 250 km/s line width as well as reasonable values for the Ne X,
Mg XI and Fe XVII lines (see Table 2 and Table 3), but it
over-predicts the oxygen equivalent widths assuming solar abundances
(including the Allende-Prieto, Lambert \& Asplund (2001) value for
oxygen).  A narrower width would saturate the oxygen lines and reduce
their equivalent widths, but predict incorrect line ratios.  In
reality it is likely that the actual absorption profiles are complex
mixtures of wide and narrow features, some of which are saturated and
some not.  See Table 3 for a list of the equivalent widths predicted
by the model as a function of velocity width.

The models calculated assuming solar abundances and listed in Table 3
indicate that non-solar abundances may be required to describe the
observed absorption spectrum.  The relative abundances of the O VIII
and Ne IX ions scale together with ionization parameter for the
ionizing spectrum determined from the X-ray spectrum.  Solar
abundances then imply that the equivalent widths of the O VIII
16.006\AA~ and Ne IX 13.447\AA~ lines should be nearly equal, while
the Ne IX line is observed to be more that three times stronger than
the O VIII line.  Similarly, the ratio of the Mg XI line to the upper
limit on Ne X is very difficult to match with standard abundance
ratios.  Table 4 shows a model with a density of $8 \times 10^{13}~\rm
cm^{-3}$ and thickness of $2 \times 10^6$ cm at $2 \times 10^{11}$~cm
from the central source, but with the Ne abundance increased by a
factor of three and the Mg abundance increased by a factor of six.
Line widths between about 150~km/s and 250~km/s match the observations
quite well.  Given the modest number of lines available and the
likelihood that the velocity structures and optical depth effects are
more complicated that we have assumed, we cannot determine abundances
precisely, but the models provide convincing evidence for a factor of
2-4 enhancement in the Ne/O ratio in the absorbing gas.

While Ne overabundances have been inferred via line ratios in at least
one other X-ray binary (the pulsar 4U~1626$-$67; Schulz et al.\ 2001),
optical observations may point to a different explanation for the
apparent overabundance in GX 339$-$4.  Analysis of the Bowen blend in
Scorpius X-1 reveals N lines which are much stronger than C lines,
indicating that CNO processing in the companion may have converted C
and O to N in this system (Steeghs \& Casares 2002).  In this
scenario, Ne might appear to be over-abundant relative to O, because O
is actually depleted due to the CNO processing.  The Bowen blend
profiles that were used to constrain the mass function of GX 339$-$4
also reveal N lines that are much stronger than C lines (Hynes et al.\
2003).  Therefore, in GX~339$-$4 it may also be the case that O is
actually depleted as a result of CNO processing in the companion star.

Any slab having the same ionization parameter ($ n \propto r^{-2}$)
and appropriate thickness ($\delta r \propto r^{2}$) would give the
same equivalent widths.  The posited slab is very thin compared to its
distance from the central source if it is located in the system itself
(that is, within a distance of a few solar radii).  If the absorbing
region is really a shell at $2\times 10^{11}$~cm, then its mass is
about $10^{-13}~M_{\odot}$.  As we expect the mass inflow rate to be
on the order of $10^{-9}~M_{\odot}$/year (Frank, King, \& Raine 2003),
such a shell could be created via a short-timescale disk wind that is
a small fraction of the mass inflow.  The shell itself may actually be
a sheet-like wind which has a small filling factor.  Moreover, the
absorber may contain dense clumps.  Certainly, the presence of the Ne
II and Ne III lines is suggestive of cold clumps a wind outflow.  If
the absorber is actually located at a radius much larger than the size
of the binary, the filling factor must be small if the mass
outflow rate implied be the absorber is to be a small fraction of the
mass inflow rate.  This is again consistent with a clumpy absorbing
geometry or a sheet-like wind outflow that has a small filling factor.
The warm absorber geometry in Seyfert 1 galaxies may not be
homogeneous, but in fact comprised of clouds which are radiatively
driven from the central accretion engine.  Indeed, mass upper limits
derived assuming radiatively-driven absorbing clouds are very similar
to estimates derived via reverberation mapping and velocity dispersion
laws for a number of Seyfert 1 galaxies (Morales and Fabian 2002).
Elvis (2000) has described a model for AGN in which winds flow from
the disk in a sheet-like geometry.

We note that the X-ray absorption might also be related to the winds
observed in cataclysmic variables (e.g. Drew 1987; Mauche \& Raymond
1987).  Those winds arise in the inner disk or boundary layer.  Their
velocities are much faster than those derived for GX~339$-$4 and XTE
J1650$-$500, typically above 2500~km/s.  Radiation pressure in
spectral lines makes a large contribution to the acceleration of CV
winds (e.g. Proga 2003), though it cannot account for all the driving
force (Mauche \& Raymond 2000). The highly ionized gas that produces
the X-ray absorption has a smaller cross section for aborbing
radiation, but the intense X-ray flux compensates and produces a high
radiation pressure.  In order to maintain the modest velocities
derived above, any wind analogous to those of CVs must arise fairly
far out in the accretion disk, or they must be clumpy enough to be
optically thick in the strong lines.

\section{Summary and Conclusions}

We have analyzed the broad-band and high-resolution X-ray spectrum of
the Galactic black hole GX~339$-$4 during outburst decline.  The
spectral resolution of the {\it Chandra}/HETGS and the high sensitivity
achieved during our long 75~ksec observation have permitted a number
of important findings.  

First, the broad Fe~K$\alpha$ line previously observed in
GX~339$-$4 (e.g., Nowak et al.\ 2002) is indeed {\it intrinsically}
broad and skewed, indicating that it is likely shaped by relativistic
effects at the inner edge of an accretion disk.  Fits with
relativistic line models and relativistically-smeared reflection
models suggest that the black hole in GX~339$-$4 may have significant
angular momentum.

Second, we have clearly detected X-ray absorption lines from He-like
and H-like O, and He-like Ne and Mg.  This is the first time that
highly ionized X-ray absorption lines from elements with $Z<20$ have
been observed in a dynamically-constrained Galactic black hole system
with a low-mass donor star.  These absorption lines may be due to
absorption in the coronal ISM, but are far more likely to be due to a
photoionized warm absorber intrinsic to the system.  The observed
blue-shift of the absorption lines indicate the absorbing material
may be outflowing from the system.  Models describing a warm absorber
with a modest ionization parameter ($\xi = 70$) and velocity width
($150-250$~km/s), and illuminated by a combination of blackbody and
power-law fluxes of the magnitude we have measured in GX~339$-$4, can
approximate the observed absorption spectrum.  The absorber may have
resulted from a disk-driven outflow, may be clumpy, and qualitatively
similar to the warm absorber geometries inferred in Seyfert 1
galaxies.

Thirdly, and most importantly, this observation has likely established
another important link between stellar-mass Galactic black holes and
supermassive black holes in AGN.  It has become clear that skewed,
relativistic Fe~K$\alpha$ lines imply that the inner accretion flow
geometry in some Galactic black hole states may be very similar to
that in Seyfert 1 galaxies (Wilms et al.\ 2001, Miller et al.\ 2002a,
Miller et al.\ 2002c).  The detection of large-scale relativistic jets
in X-rays in XTE~J1550$-$564 (Corbel et al.\ 2002) may provide a
window into large-scale jet formation in some AGN.  This observation
indicates that X-ray warm absorbers --- plausibly produced via a disk
wind --- may be common to both stellar mass black holes and Seyfert
galaxies.  The long dynamical timescales in AGN make it difficult to
observe changes in the warm absorber and to clearly link effects to
the central accretion engine.  Future sensitive observations of
Galactic black holes may reveal that stellar-mass black holes are
accessible laboratories for studying the nature of X-ray warm
absorbers. 

\section{Acknowledgments}
We acknowledge the anonymous referee for helpful suggestions which
have improved the clarity and content of this paper.  We wish to thank
CXC Director Harvey Tananbaum and the CXC staff for granting this DDT
observation.  JMM thanks the University of Cambridge Institute of
Astronomy for its hospitality, and Roderick Johnstone for great
computing support.  We wish to thank Danny Steeghs, Adrian Turner,
Anil Pradhan, Randall Smith, and Norbert Schulz for useful
discussions, and John Houck for assistance with ISIS.  JMM gratefully
acknowledges support from the NSF through its Astronomy and
Astrophysics Postdoctoral Fellowship program.  JAT acknowledges
partial support from NASA grant NAG5--13055.  WHGL gratefully
acknowledges support from NASA.  This research has made use of the
data and resources obtained through the HEASARC on-line service,
provided by NASA-GSFC.

\clearpage

\centerline{~\psfig{file=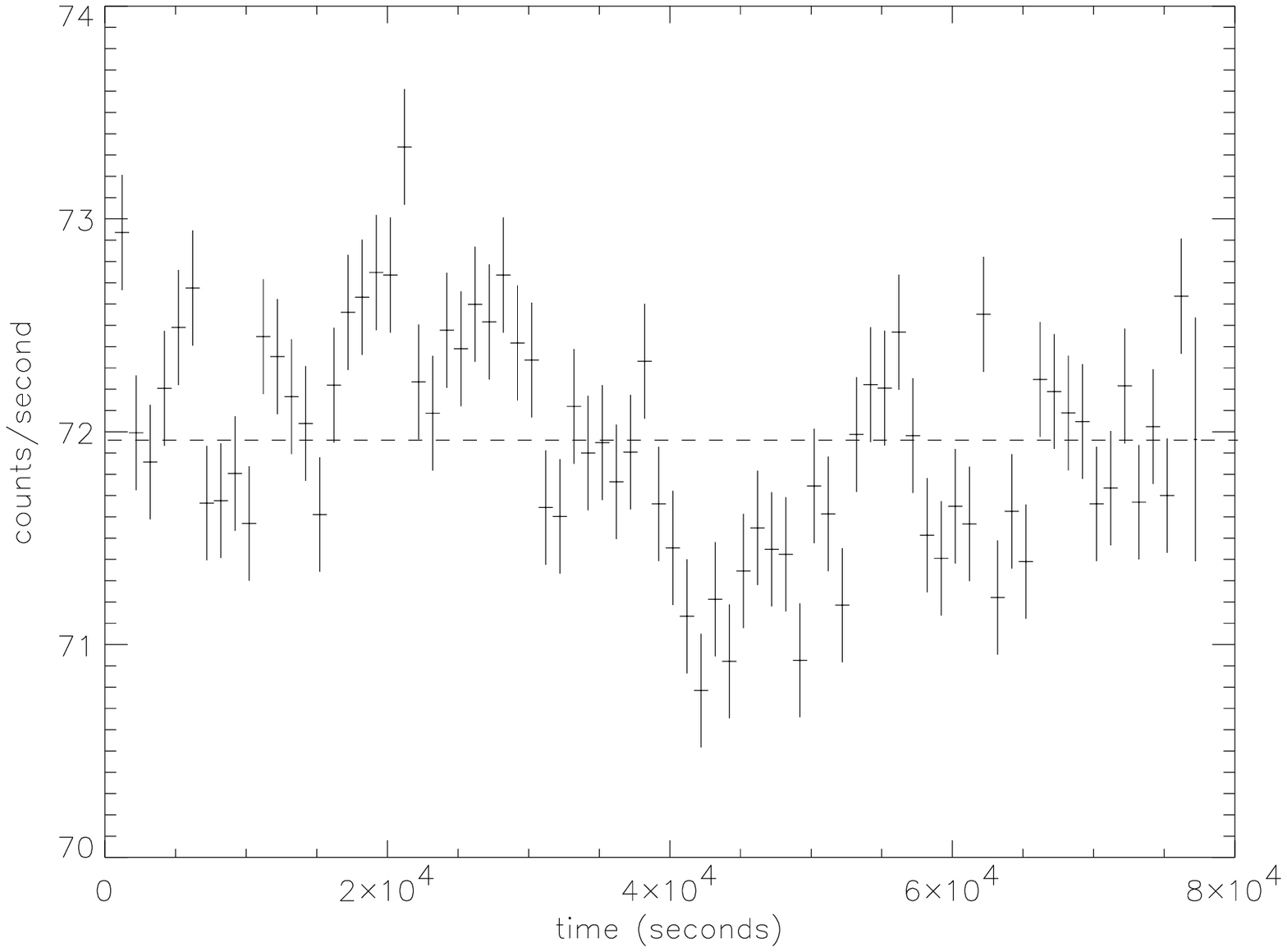,width=4.5in}~}
\figcaption[h]{\footnotesize {\it Chandra} observed GX 339-4 for
  approximately 75~ksec starting on 2003 March 17.8.  The
  0.3--10.0~keV dispersed counts (summed HEG and MEG first order)
  lightcurve is shown above.  Each time bin is 1000 seconds.  The
  dashed horizontal line indicates the mean count rate.  The flux is
  very steady over the course of the observation; variations are less
  than 2\% of the mean.}
\medskip

\clearpage

\begin{table}[t]
\caption{Models for the 1.2--10.0~keV Spectrum of GX 339$-$4 and XTE~J1650$-$500}
\begin{footnotesize}
\begin{center}
\begin{tabular}{llllllll}
\multicolumn{3}{l}{Observation} & GX 339$-$4 & GX~339$-$4 & GX~339$-$4 & XTE J1650$-$500 A & XTE J1650$-$500 B \\
\multicolumn{3}{l}{~} & (simple model) & (pexriv$^{a}$) & (CDID$^{b}$) & (simple model)$^{c}$ & (simple model)$^{c}$\\
\tableline
\multicolumn{3}{l}{$N_{H}~(10^{21}~{\rm cm}^{-2})$} & 3.9(1) & 4.3(1) & 4.3(2) & 5.6(2) & 5.3(1) \\
\tableline
\multicolumn{3}{l}{\bf MCD} & ~ & ~ & ~ & ~ & ~ \\
\multicolumn{3}{l}{$kT$ (keV)} & 0.569(4) & 0.56(1) & 0.56(1) & 0.610(4) & 0.570(1) \\
\multicolumn{3}{l}{Norm. ($10^{3}$)} & 2.04(5) & 2.15(5) & 2.21(5) & 10.3(2) & 10.6(2) \\
\tableline
\multicolumn{3}{l}{\bf power-law} & ~ & ~ & ~ & ~ & ~ \\
\multicolumn{3}{l}{$\Gamma$} & 2.5(1) & 2.70$_{-0.1}$ & -- & $2.5^{+0.1}_{-0.2}$ & 5.4 \\
\multicolumn{3}{l}{Norm.} & 0.15(2) & 0.16(6) & -- & $1.8_{-0.6}^{+0.3}$ & $<0.1$ \\
\tableline
\multicolumn{3}{l}{\bf Laor/smearing} & ~ & ~ & ~ & ~ & ~ \\
\multicolumn{3}{l}{E (keV)} & 6.8(2) & $6.97_{-0.05}$ & -- & -- & -- \\
\multicolumn{3}{l}{$R_{in}~(R_{g})$} & $2.5^{+2.0}_{-0.3}$ & $2.0_{-0.7}^{+2.2}$ & $1.3^{+1.7}_{-0.1}$ & -- & -- \\
\multicolumn{3}{l}{$q$} & 4.5(6) & $3.0(2)$ & $3.0^{+0.7}_{-0.4}$ & -- & -- \\
\multicolumn{3}{l}{$i$} & 40(20) & $15^{+15}$ & 30(15) & -- & -- \\
\multicolumn{3}{l}{Norm. ($10^{-3}$)} & $1.8^{+0.5}_{-0.3}$ & $0.9(4)$ & -- & -- & -- \\ 
\multicolumn{3}{l}{$EW$ (eV)} & $600^{+200}_{-100}$ & $770^{+310}_{-310}$ & -- & -- & -- \\ 
\tableline
\multicolumn{3}{l}{\bf reflection} & ~ & ~ & ~ & ~ & ~ \\
\multicolumn{3}{l}{$\Gamma$} & -- & -- & 2.6(1) & -- & -- \\
\multicolumn{3}{l}{R} & -- & 0.9(4) & $1.2^{2.0}_{-0.5}$ & -- & -- \\
\multicolumn{3}{l}{E$_{cut}$ (keV)} & -- & 200 & -- & -- & -- \\
\multicolumn{3}{l}{$T_{disk}~(10^{7}~{\rm K})$} & -- & 1.2 & -- & -- & -- \\
\multicolumn{3}{l}{log($\xi$)~($10^{4}$)} & -- & 4.0 & 4.0(4) & -- & -- \\
\multicolumn{3}{l}{Norm.} & -- & -- & $7(3) \times 10^{-27}$ & -- & -- \\
\tableline
\multicolumn{3}{l}{$\chi^{2}/\nu$} & 4116.2/3625$^{\dag}$ & 4024.2/3619 & 4069.9/3619 & 4411.8/3631 & 4656.3/3629 \\
\tableline
\multicolumn{3}{l}{$f_{hard, 0.5-10}$}  & 0.13 & 0.18 & 0.26 & 0.21 & $<0.03$ \\
\multicolumn{3}{l}{F$_{0.5-10}$~($10^{-8}$~cgs)} & 0.42(2) & 0.44(2) & 0.48(2) & 3.1(1) & 1.9(1) \\
\multicolumn{3}{l}{L$_{0.5-10}$~($10^{37}$~erg/s, $d=4$kpc)} & 0.80(4) & 0.84(4) & 0.9(4) & 5.9(2) & 3.6(1) \\ 
\multicolumn{3}{l}{L$_{0.5-10}$~($10^{37}$~erg/s, $d=5$kpc)} & 1.26(6) & 1.32(6) & 1.44(6) & 9.3(3) & 5.7(2) \\ 
\tableline
\end{tabular}
\vspace*{\baselineskip}~\\ \end{center} 
\tablecomments{The results of fits to the combined first-order spectra
  of GX 339$-$4 and XTE~J1650$-$500 in the 1.2--10.0~keV band are
  listed above.  Errors are 90\% confidence errors.  Where errors are
  not quoted, the parameter was fixed at the value reported.  The
  power-law indices were constrained to be within $\Delta{\Gamma} \leq
  0.2$ of the value measured with {\it RXTE} in the 3--200~keV band.
  An ``inverse'' edge component (E$=$2.07~keV, $\tau=-0.13$) was
  included in all models to account for the well-known instrumental Ir
  edge.  The fluxes quoted above are ``unabsorbed'' fluxes.  The fits
  obtained are not formally acceptable; the poor reduced $\chi^{2}$
  values very likely reflect calibration problems (such as the Ir
  edge) rather than any deficiency with the models used.\\ $^{a}$ The
  reflection model was ``smeared'' or convolved with the Laor line
  element, with all parameters of the smearing function and separate
  Laor line component itself constrained to be equivalent.  The
  power-law index and normalization within the smeared ``pure''
  reflection component were fixed to those of the external
  power-law.\\ $^{b}$ the reflection model was ``smeared'' with the
  Laor line component; the reflection model includes line emission
  explicitly.\\ $^{c}$ We have only fit simple models to XTE
  J1650$-$500; the position of chip edges due to the chosen Y offset
  complicates detection of broad Fe~K$\alpha$ lines revealed in
  simultaneous {\it RXTE} data.\\ $^{\dag}$ The best fit model without
  the Laor component gives $\chi^{2}/\nu = 4176.5/3630$, or an
  F-statistic of $3.6 \times 10^{-10}$, signaling this component is
  required at the $6.3\sigma$ level of confidence.}
\vspace{-1.0\baselineskip}
\end{footnotesize}
\end{table}

\clearpage

\centerline{~\psfig{file=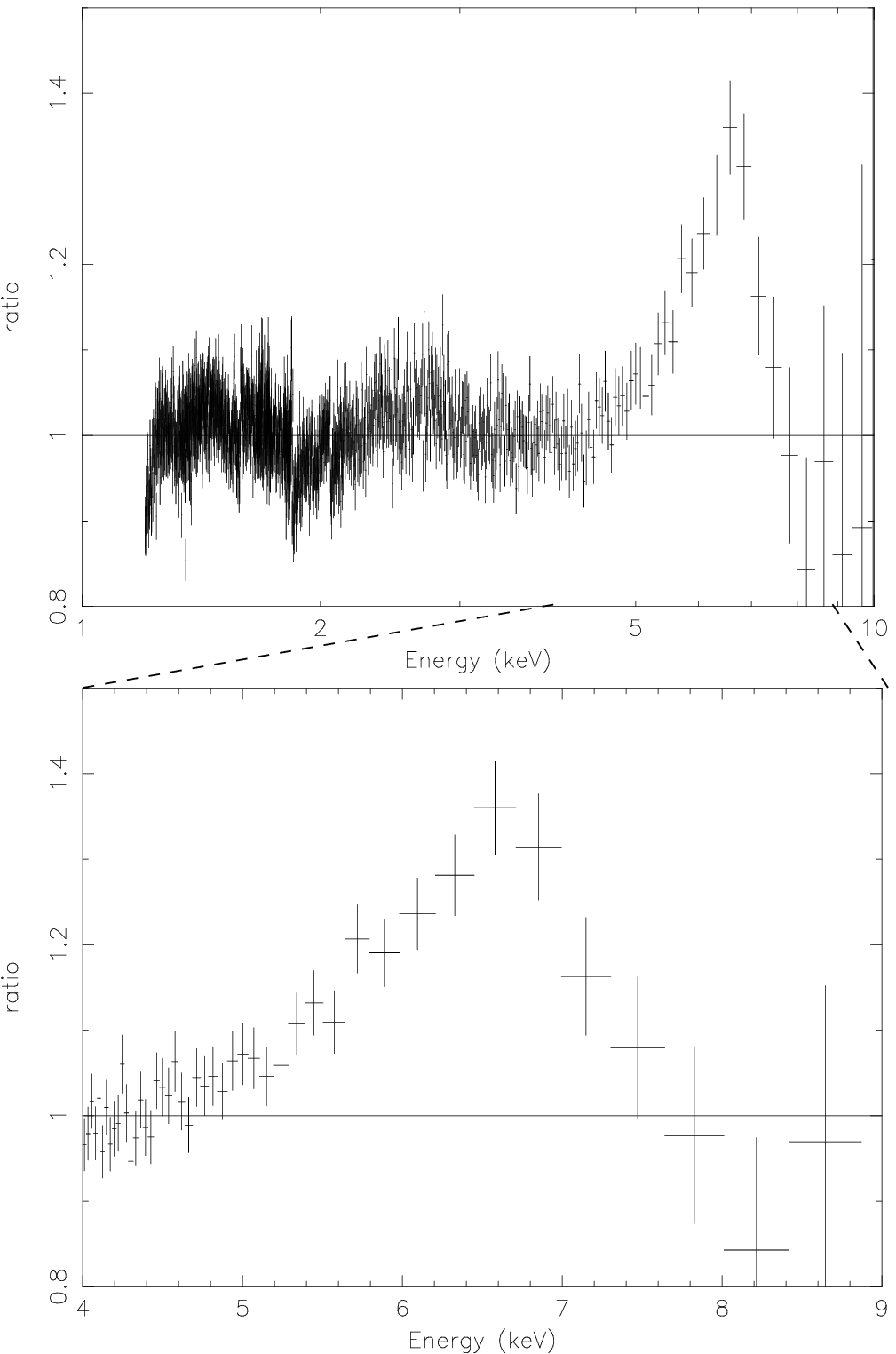,width=4.5in}~}
\figcaption[h]{\footnotesize The data/model ratio obtained when the
  time-averaged combined first-order HEG spectra from GX~339$-$4 are
  fit with an MCD plus power-law model, assuming $\Gamma=2.5$ (as per
  fits to simultaneous {\it RXTE} observations on the 3.0--100.0~keV
  band).  The 4.0--7.0~keV line region was ignored in fitting the
  model, and the data have been rebinned for visual clarity.  An
  asymmetric Fe~K$\alpha$ emission line is clearly revealed.  The
  broad line profile signals that the disk extends close to the black
  hole even at the mass accretion rate ($\dot{m}_{Edd.}  \simeq 0.1$)
  inferred from this observation.}
\medskip

\clearpage

\centerline{~\psfig{file=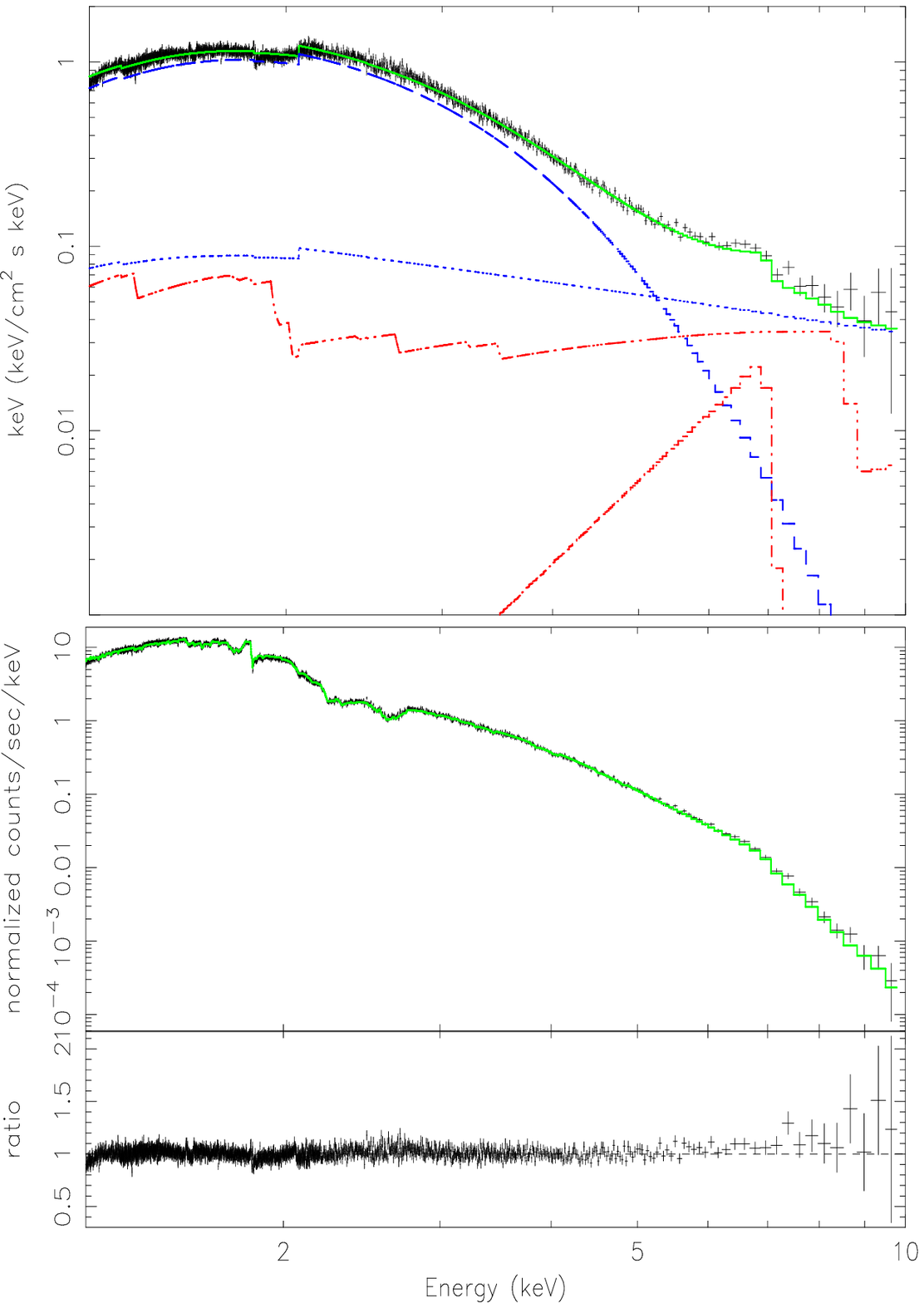,width=4.5in}~}
\figcaption[h]{\footnotesize The combined first-order HEG spectra of
  GX~339$-$4 fit with a relativistically-smeared reflection model
  (``pexriv'', see Table 1) in
  the 1.2--10.0~keV band.  The plot above shows the ``unfolded''
  spectrum and components within the model.  The total model is shown
  in green, the MCD and power-law continuum are shown in blue, and the
reflection model and Laor Fe~K$\alpha$ emission line component are
shown in red.  The reflection model is shown prior to smearing.  Below, the count spectrum and data/model ratio are
shown, with the model again shown in green.  In both plots, the
spectrum has been rebinned for visual clarity.}
\medskip

\clearpage

\centerline{~\psfig{file=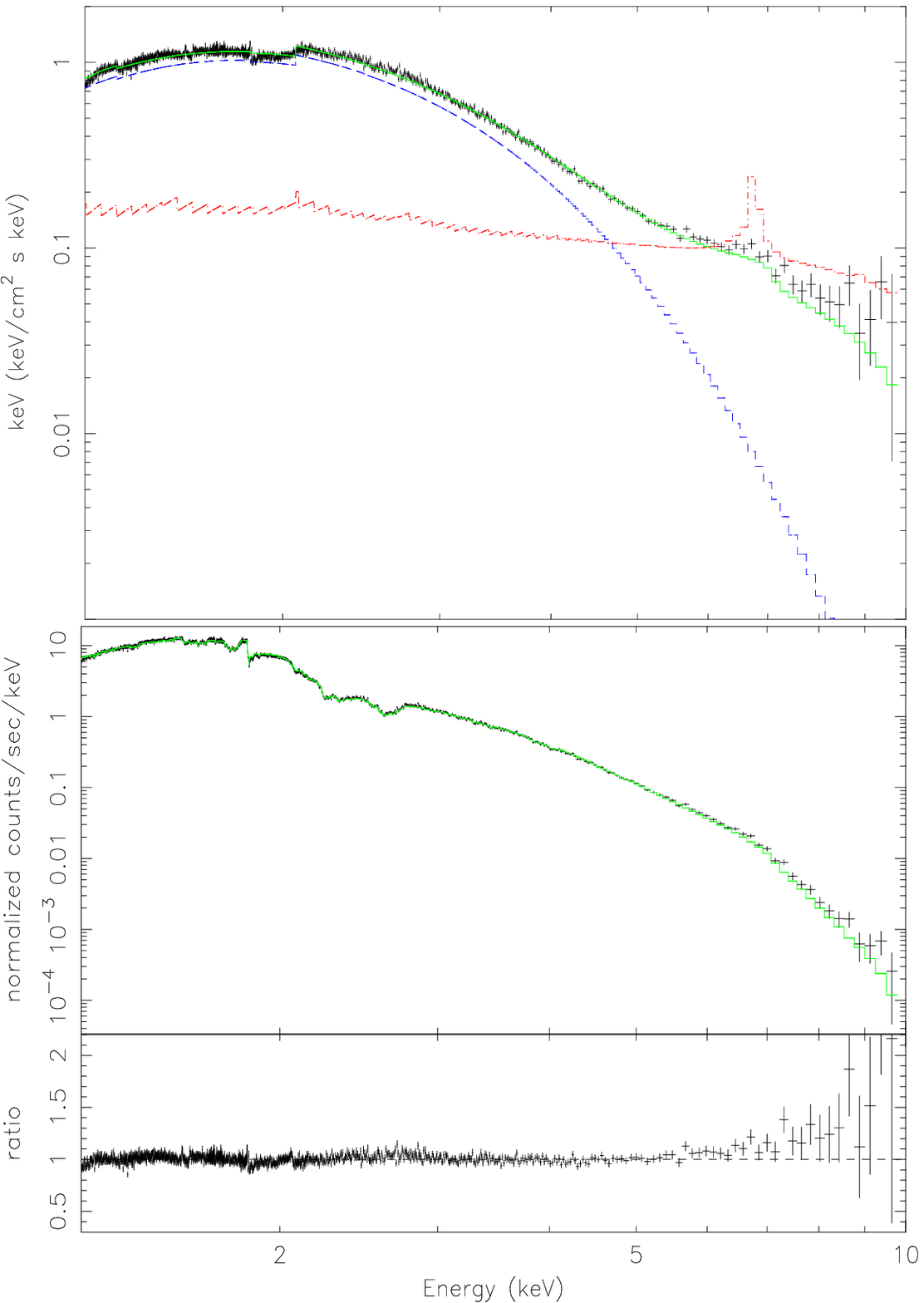,width=4.5in}~}
\figcaption[h]{\footnotesize The combined first-order HEG spectra of
  GX~339$-$4 fit with a relativistically-smeared reflection model
  (``CDID'', see Table 1) in the 1.2--10.0~keV band.  The plot above
  shows the ``unfolded'' spectrum and components within the model.
  The total model is shown in green, the MCD component in blue, and
  the reflection model (which includes an Fe~K$\alpha$ emission line
  component, and is shown prior to smearing) are shown in red.  Below,
  the count spectrum and data/model ratio are shown, with the model
  again shown in green.  In both plots, the spectrum has been rebinned
  for visual clarity.}
\medskip

\clearpage

\begin{table}[t]
\caption{Absorption Lines and Line Flux Limits in GX 339-4 and XTE J1650$-$500}
\begin{footnotesize}
\begin{center}
\begin{tabular}{lllllllllll}
\multicolumn{3}{l}{Ion and} & Theor. & Meas. & Shift & \multicolumn{2}{c}{FWHM} & W & Flux & N$_{\rm Z}$ \\
\multicolumn{3}{l}{Transition} & (\AA) & (\AA) & (km/s) & ($10^{-3}$\AA) & (km/s) & (m\AA) & ($10^{-4}$~ph/cm$^{2}$/s) & ($10^{16}~{\rm cm}^{-2}$) \\
\tableline

\multicolumn{11}{c}{--- GX~339$-$4 ---}\\

\tableline

\multicolumn{3}{l}{O VIII $1s-2p$} & 18.967 & 18.964(1) & $50^{+20}_{-20}$ & $23_{-22}^{+11}$ & $370_{-350}^{+170}$ & $-19(5)$ & $-0.8(2)$ & $1.5(4)$\\

\multicolumn{3}{l}{O VII $1s^{2}-1s3p$} & 18.627 & 18.623(1) & $60^{+20}_{-10}$ & $9_{-9}^{+18}$ & $140_{-140}^{+280}$ & $-14^{-5}_{+3}$ & $-0.7(2)$ &
$3.2^{+1.0}_{-0.7}$\\

\multicolumn{3}{l}{O VIII $1s-3p$} & 16.006 & 16.003(4) & $60^{+70}_{-80}$ & $23_{-23}^{+19}$ & $430_{-430}^{+360}$ &
$-5(2)$ & $-0.6^{-0.3}_{+0.2}$ & $3(1)$\\

\multicolumn{3}{l}{Ne II} & 14.631 & 14.606(1) & 510(20) & $20^{+6}_{-5}$ & $400^{+130}_{-100}$ &
$-13(1)$ & $-2.7(3)$ & 11(1)\\

\multicolumn{3}{l}{Ne III} & 14.526 & 14.504(2) & 450(40) & $24^{+5}_{-5}$ & $500^{+100}_{-100}$ &
$-11(1)$ & $-2.4(3)$ & 5.7(7)\\

\multicolumn{3}{l}{Ne IX $1s^{2}-1s2p$} & 13.447 & 13.442(1) & $110^{+30}_{-20}$ & $16(4)$ & $350^{+80}_{-90}$ & $-17(1)$ & $-3.8(3)$ & $1.4(1)$\\

\multicolumn{3}{l}{Ne IX $1s^{2}-1s3p$} & 11.547 & $11.541^{+0.004}_{-0.001}$ & $160^{+20}_{-100}$ & $3.0^{+10}_{-3}$ & $80^{+250}_{-80}$ & $-3(1)$ & $-1.2(1)$ & $1.7(6)$\\

\multicolumn{3}{l}{Ne IX $1s^{2}-1s4p$} & 11.000 & 10.998(3) & $60^{+80}_{-90}$ & $2.4^{+8.2}_{-2.4}$ & $70^{+240}_{-70}$ & $-1.6^{-0.2}_{+0.4}$ & $-0.8^{-0.1}_{+0.2}$ & $2.0^{+0.5}_{-0.3}$\\

\multicolumn{3}{l}{Mg XI $1s^{2}-1s2p$} & 9.169 & $9.166^{+0.001}_{-0.002}$ & $100^{+60}_{-30}$ & $10_{-8}^{+5}$ & $330_{-250}^{+170}$ & $-2.4^{-0.2}_{+0.1}$ & $-1.5^{-0.1}_{+0.2}$ & $0.45^{+0.03}_{-0.02}$\\

\tableline

\multicolumn{3}{l}{O VII $1s^{2}-1s4p$} & 17.768 & -- & -- & $24$ & $400$ & $-6^{-3}_{+3}$ & $-0.4(2)$ & $4(2)$\\

\multicolumn{3}{l}{O VIII $1s-4p$} & 15.176 & -- & -- & $20$ & $400$ & $-2.2^{-0.5}_{+1.0}$ & $-0.4^{-0.1}_{+0.2}$ & $4^{+1}_{-2}$\\

\multicolumn{3}{l}{Fe XVI} & 16.63 & -- & -- & $22$ & $400$ & $-1^{-2}_{+1}$ & $-0.1^{-0.2}_{+0.1}$ & -- \\

\multicolumn{3}{l}{Fe XVII} & 15.015 & -- & -- & $20$ & $400$ & $-1.6^{-1.6}_{+1.0}$ & $-0.3^{-0.3}_{+0.2}$ & $0.03^{+0.03}_{-0.02}$\\

\multicolumn{3}{l}{Fe XVII} & 13.823 & -- & -- & $18$ & $400$ & $-0.7^{-1.1}_{+0.7}$ & $-0.2^{-0.3}_{+0.2}$ & $0.10^{+0.15}_{-0.10}$\\

\multicolumn{3}{l}{Ne X $1s-2p$} & 12.1339 & -- & -- & $16$ & $400$ & $-0.6(6)$ & $-0.2(2)$ & $0.1(1)$\\

\multicolumn{3}{l}{Fe XVII} & 12.120 & -- & -- & $16$ & $400$ & $-0.6^{-0.9}_{+0.6}$ & $-0.2^{-0.3}_{+0.2}$ & $0.06^{+0.09}_{-0.06}$\\

\multicolumn{3}{l}{Mg XII $1s-2p$} & 8.421 & -- & -- & $11$ & $400$ & $-0.06(6)$ & $-0.04(4)$ & $0.03(3)$\\

\multicolumn{3}{l}{Si XII $1s^{2}-1s2p$} & 6.648 & -- & -- & $8.5$ & $400$ & $-9(5)$ & $-0.6(3)$ & $0.3(2)$\\

\multicolumn{3}{l}{Si XIV $1s-2p$} & 6.61822 & -- & -- & $8$ & $400$ & $-0.4(4)$ & $-0.3(3)$ & $0.3(3)$\\

\tableline

\multicolumn{11}{c}{--- XTE J1650$-$500 Observation A ---}\\

\tableline

\multicolumn{3}{l}{Ne II} & 14.631 & 14.606(3) & 510(60) & $26^{+8}_{-7}$ & $530^{+160}_{-140}$ &
$-8(1)$ & $-9(2)$ & $7(1)$\\

\multicolumn{3}{l}{Ne III} & 14.526 & -- & -- & -- & -- & $-4(1)$ & $-4(1)$ & $2.0(6)$\\

\multicolumn{3}{l}{Ne IX $1s^{2}-1s2p$} & 13.447 & -- & -- & -- & -- & $-0.7^{-1.1}_{+0.7}$ & $-0.8^{-1.4}_{+0.8}$ & $0.06^{+0.09}_{-0.06}$\\

\multicolumn{3}{l}{Mg XI $1s^{2}-1s2p$} & 9.169 & -- & -- & -- & -- & $-0.5^{-0.3}_{+0.5}$ & $-2^{-1}_{+2}$ & $0.09^{+0.05}_{-0.09}$\\

\tableline

\multicolumn{11}{c}{--- XTE J1650$-$500 Observation B ---}\\

\tableline

\multicolumn{3}{l}{Ne II} & 14.631 & 14.609(6) & 450(90) & $12^{+8}_{-12}$ & $240^{+160}_{-240}$ &
$-6(1)$ & $-4.7(8)$ & 5.2(9)\\

\multicolumn{3}{l}{Ne III} & 14.526 & 14.510(5) & -- & $2.4^{+2.4}_{-2.4}$ & $50^{+50}_{-50}$ &
$-2.6(8)$ & $-2.1(7)$ & 1.4(5)\\

\multicolumn{3}{l}{Ne IX $1s^{2}-1s2p$} & 13.447 & 13.450(3) & $-70^{-60}_{+70}$ & $16(4)$ & $2.4^{+2.4}_{-2.4}$ & $-3.9(9)$ & $-3.5(8)$ & $0.34(8)$\\

\multicolumn{3}{l}{Mg XI $1s^{2}-1s2p$} & 9.169 & -- & -- & -- & -- & $-0.7^{-2.5}_{+0.7}$ & $-0.2^{-0.7}_{+0.2}$ & $0.1^{+0.5}_{-0.1}$\\

\tableline
\end{tabular}
\vspace*{\baselineskip}~\\ \end{center} 
\tablecomments{Absorption lines found in the {\it Chandra}/HETGS MEG
  spectra of GX 339$-$4 and XTE J1650$-$500.  The continua were fit
  locally using power-law models modified by neutral photoelectric
  absorption edges (due to the ISM) where appropriate.  The lines were
  fit with simple Gaussian models (see Figures 5, 6, and 6).  The
  errors quoted above are $1\sigma$ uncertainties.  Line widths
  consisent with zero are not unresolved.  For GX 339$-$4, we have
  included upper-limits on important lines, as these limits which may
  enable others to readily develop models for the nature of the
  absorber.  Where an XTE J1650$-$500 line wavelength measurement is
  not given, the line parameters are upper-limits assuming the same
  FWHM and centroid wavelength as the corresponding GX 339$-$4 line.
  Line wavelengths and oscillator strengths are taken from Verner,
  Verner, and Ferland (1996), except parameters for Ne~II and Ne~III
  which are taken from Behar and Netzer (2002).}
\vspace{-1.0\baselineskip}
\end{footnotesize}
\end{table}

\vspace{5.0in}

\clearpage

\begin{table}[t]
\caption{Photoionized Plasma Models for GX~339$-$4 Assuming Solar Abundances}
\begin{footnotesize}
\begin{center}
\begin{tabular}{llllllllll}
\multicolumn{2}{l}{Ion \& Wavelength (\AA)} & \multicolumn{2}{l}{EW (Meas., m\AA)} & EW (m\AA) & EW (m\AA) & EW (m\AA) & EW (m\AA) & EW (m\AA) & EW (m\AA) \\
\tableline
\multicolumn{2}{l}{~} &  \multicolumn{2}{l}{~} & 50~km/s & 150~km/s &250~km/s & 350~km/s &  450~km/s & 550~km/s \\
\tableline
 
\multicolumn{2}{l}{O VIII 18.967\AA} &  \multicolumn{2}{l}{19.0} & 11.63 & 28.58 &  41.36 & 50.85 & 57.32 &  64.51 \\

\multicolumn{2}{l}{O VII  18.627\AA} &  \multicolumn{2}{l}{14.0} & 7.38 &  12.24 &  13.84 & 14.64 & 15.12 &  15.45 \\

\multicolumn{2}{l}{O VIII 16.006\AA} &  \multicolumn{2}{l}{5.0} & 6.41 & 10.65 &  12.06 &  12.77 &  13.20 &  13.51 \\

\multicolumn{2}{l}{Ne IX 13.447\AA} & \multicolumn{2}{l}{17.0} & 5.86 & 10.36 &  12.05 &  12.88 &  13.44 &  13.79 \\

\multicolumn{2}{l}{Ne IX 11.547\AA} &   \multicolumn{2}{l}{3.0} &  1.88 & 2.18 &  2.25  &  2.30 &  2.30 &  2.30 \\

\multicolumn{2}{l}{Ne IX 11.000\AA} & \multicolumn{2}{l}{1.6} & 0.81 & 0.84 & 0.86 & 0.86 & 0.86 & 0.8 \\

\multicolumn{2}{l}{Ne X 12.13\AA}  &  \multicolumn{2}{l}{$<3.0$} &  3.00 & 3.85 &  4.06  &  4.15 &  4.21 &  4.26 \\

\multicolumn{2}{l}{Mg XI  9.166\AA} &  \multicolumn{2}{l}{2.4} &  1.39 & 1.60 &  1.64  &  1.67  &  1.67 &  1.67 \\
 
\multicolumn{2}{l}{Fe XVII  15.010\AA} & \multicolumn{2}{l}{$<3.0$} & 5.03 &  7.38 &  8.08 &  8.43  & 8.59 & 8.72 \\

\multicolumn{2}{l}{Fe XVIII 14.610\AA} &  \multicolumn{2}{l}{$<10.0$} & 0.31 & 0.31 &  0.31 &  0.31  &  0.31 &  0.31 \\

\tableline
\end{tabular}
\vspace*{\baselineskip}~\\ \end{center} 
\tablecomments{Absorption line equivalent widths (multiplied by $-1$)
  as a function of velocity width for a photoionized absorption models
  based on the code of Raymond (1993); see Section 4.4 for details.
  The models above were generated assuming solar elemental
  abundances.  A velocity width of 150~km/s is likely the best match
  to the absorption lines observed in GX~339$-$4; abundances differing
  slightly from solar values may be required to explain the observed
  Mg/O and Ne/O ratios (see Section 4.4).}
\vspace{-1.0\baselineskip}
\end{footnotesize}
\end{table}

\vspace{5.0in}

\clearpage

\begin{table}[t]
\caption{Photoionized Plasma Models for GX~339$-$4 with Non-solar Abundances}
\begin{footnotesize}
\begin{center}
\begin{tabular}{llllllllll}
\multicolumn{2}{l}{Ion \& Wavelength (\AA)} & \multicolumn{2}{l}{EW (Meas., m\AA)} & EW (m\AA) & EW (m\AA) & EW (m\AA) & EW (m\AA) & EW (m\AA) & EW (m\AA) \\
\tableline
\multicolumn{2}{l}{~} &  \multicolumn{2}{l}{~} & 50~km/s & 100~km/s &150~km/s & 200~km/s &  250~km/s & 300~km/s \\
\tableline
 
\multicolumn{2}{l}{O VIII 18.967\AA} &  \multicolumn{2}{l}{19.0} & 9.72 &  16.04 & 20.26 &  23.57 & 25.94 & 27.72\\

\multicolumn{2}{l}{O VII 18.627\AA} &  \multicolumn{2}{l}{14.0} & 7.71 & 11.24 & 13.12 &  14.24 & 14.98 & 15.52\\

\multicolumn{2}{l}{O VIII 16.006\AA} &   \multicolumn{2}{l}{5.0} & 3.79  &  4.49 & 4.79 &  4.95 &  5.03 &  5.10\\

\multicolumn{2}{l}{Ne IX 13.447\AA} & \multicolumn{2}{l}{17.0} &  6.47 &  10.11 & 12.75 &  14.42 & 15.56 & 16.42\\

\multicolumn{2}{l}{Ne IX 11.547\AA} &  \multicolumn{2}{l}{3.0} &  2.46 &  2.85 &  3.01 & 3.09 &  3.14 & 3.17\\

\multicolumn{2}{l}{Ne IX 11.000\AA} &   \multicolumn{2}{l}{1.6} &  1.12 &  1.19 &  1.21 & 1.23 &  1.23 &  1.23\\

\multicolumn{2}{l}{Ne X 12.13\AA} &   \multicolumn{2}{l}{$<3.0$} &  1.72 &  1.90 &  1.96 &  1.99 &  2.01 &  2.04\\

\multicolumn{2}{l}{Mg XI 9.166\AA} &   \multicolumn{2}{l}{2.4} &  1.40 &  1.56 &  1.61 &  1.64 &  1.66 &  1.68\\

\multicolumn{2}{l}{Fe XVII 15.010\AA} &  \multicolumn{2}{l}{$<3.0$} &  0.67 & 0.69 &  0.69 &  0.69 &  0.69 &  0.69\\

\multicolumn{2}{l}{Fe XVIII 14.610\AA} &  \multicolumn{2}{l}{$<10.0$} &  0.02 & 0.02 &  0.02 &  0.02 &  0.02  & 0.02\\

\tableline
\end{tabular}
\vspace*{\baselineskip}~\\ \end{center} 
\tablecomments{Absorption line equivalent widths (multiplied by $-1$)
  as a function of velocity width for a photoionized absorption models
  based on the code of Raymond (1993); see Section 4.4 for details.
  The models above were generated assuming solar elemental
  abundances for all elements except Ne (three times solar) and Mg
  (six times solar).  Velocity widths in the 150--250~km/s range match
  the absorption lines observed in GX~339$-$4 rather well.}
\vspace{-1.0\baselineskip}
\end{footnotesize}
\end{table}

\clearpage

\centerline{~\psfig{file=f5.ps,width=4.5in}~}
\centerline{~\psfig{file=f6.ps,width=4.5in}~}
\centerline{~\psfig{file=f7.ps,width=4.5in}~}
\figcaption[h]{\footnotesize Slices of the combined first-order MEG
  spectra from GX 339-4, fit with simple power-law continua (modified
  by neutral ISM absorption edges where appropriate) and Gaussian
  models for the Ne and Mg absorption lines.  The best-fit model is
  shown in red, and 1$\sigma$ error bars are shown in blue.  See Table
  2 for line identifications and properties, and see Tables 3 and 4
  for line equivalent widths predicted by photoionzed absorber models
  with varying velocity width lines.}

\centerline{~\psfig{file=f8.ps,width=4.5in}~}
\centerline{~\psfig{file=f9.ps,width=4.5in}~}
\figcaption[h]{\footnotesize Slices of the combined first-order MEG
  spectra from GX 339-4, fit with simple power-law continua (modified
  by neutral ISM absorption edges where appropriate) and Gaussian
  models for the H-like and He-like O absorption lines.  The best-fit
  model is shown in red, and 1$\sigma$ error bars are shown in blue.
  See Table 2 for line identifications and properties, and see Tables
  3 and 4 for line equivalent widths predicted by photoionzed absorber
  models with varying velocity width lines.}

\pagebreak

\centerline{~\psfig{file=f10.ps,width=4.5in}~}
\centerline{~\psfig{file=f11.ps,width=4.5in}~}
\figcaption[h]{\footnotesize The Galactic black hole candidate XTE
  J1650$-$500 was observed on two occasions (30~ksec each) with the
  {\it Chandra}/HETGS.  A slice of the MEG spectrum from the first
  observation is shown above, and the corresponding slice of the
  second observation is shown below.  The spectra were fit with simple
  power-law models, modified by neutral ISM absorption edges and
  Gaussian line models.  Note that the Ne IX line is not detected in
  the first observation, but that the Ne~II line is quite strong.  In
  contrast, the Ne IX line is detected in the second observation.  The
  Ne~II line is marginally stronger in the first observation; the
  Ne~III resonance line is not clearly detected in either spectrum.
  The first observation occurred at a source flux of roughly 0.40~Crab
  and the second at a flux of roughly 0.25~Crab, 25 days after the
  first.  This indicates that the absorbing geometry changed with
  time, linking the absorbing geometry to the accreting source instead
  of coronal gas in the ISM.}

\end{document}